\documentclass[aps,amssymb,prb,twocolumn]{revtex4} \fussy
\usepackage{ifpdf} %definieert de switch \ifpdf waarmee je pdf(la)tex  detecteert
\ifpdf
\usepackage[pdftex]{graphicx,color} % voor het verwerken van plaatjes met pdflatex of pdftex
\usepackage{epstopdf}         %roept epstopdf.exe aan, N.B. op www.CTAN.org staat dit bestand, als het niet al in C:\texmf\miktex\bin\ staat
\else

\fi

\usepackage{epsfig,graphicx}

\usepackage{float}

\begin{document}

\title{Strongly nonexponential time-resolved fluorescence of quantum-dot ensembles
in three-dimensional photonic crystals}

\author{Ivan S. Nikolaev$^{1,2}$}

\author{Peter Lodahl$^{2,3}$}

\author{A. Floris van Driel$^4$}

\author{A. Femius Koenderink$^{1}$}

\author{Willem L. Vos$^{1,2}$}
\email{w.l.vos@amolf.nl, www.photonicbandgaps.com}

\affiliation{$^1$FOM Institute for Atomic and Molecular Physics
(AMOLF), Amsterdam, The Netherlands}

\affiliation{$^2$Complex Photonic Systems, MESA$^+$ Research
Institute, University of Twente, The Netherlands}

\affiliation{$^3$COM{\Large$\cdot$}DTU, Department of
Communications, Optics, and Materials, Nano{\Large$\cdot$}DTU,
Technical University of Denmark, Kgs. Lyngby, Denmark}

\affiliation{$^4$Condensed Matter and Interfaces, Debye Institute,
Utrecht University, The Netherlands}

\begin{abstract}

We observe experimentally that ensembles of quantum dots in
three-dimensional (3D) photonic crystals reveal strongly
nonexponential time-resolved emission. These complex emission decay
curves are analyzed with a continuous distribution of decay rates.
The log-normal distribution describes the decays well for all
studied lattice parameters. The distribution width is identified
with variations of the radiative emission rates of quantum dots with
various positions and dipole orientations in the unit cell. We find
a striking sixfold change of the width of the distribution by
varying the lattice parameter. This interpretation qualitatively
agrees with the calculations of the 3D projected local density of
states. We therefore conclude that fluorescence decay of ensembles
of quantum dots is highly nonexponential to an extent that is
controlled by photonic crystals.

\end{abstract}

%\pacs{...}

\maketitle

\section{Introduction}

Control over spontaneous emission from ensembles of excited light
sources is of great interest for many applications, such as
miniature lasers, light-emitting diodes,\cite{Yab87,Park04} and
solar cells.\cite{Grataetzel01} The rate of spontaneous emission is
determined not only by the internal nature of emitters but also by
their environment.\cite{Drexhage70,Kleppner81} According to Fermi's
golden rule, this rate is proportional to the local radiative
density of optical states (LDOS), to which the emitters
couple.\cite{Sprik96,Vats02} This projected LDOS counts, at given
frequency and orientation of the transition dipoles, the number of
electromagnetic states at the locations of the emitters.
It has been predicted that periodic dielectric structures, known as
photonic crystals, can be used to radically change the
LDOS.\cite{Yab87} The main research goal is the achievement of a
photonic band gap, \emph{i.e.}, a range of frequencies where no
electromagnetic states exist inside the crystal, irrespective of the
location. It has also been predicted that a much weaker requirement
than a gap suffices to suppress spontaneous emission:\cite{Sprik96}
by placing sources at judicious locations in the crystal unit cell
where the LDOS vanishes. Since the frequency-integrated number of
states is conserved, one expects the LDOS to be strongly increased
at some frequencies outside such a
pseudogap.\cite{Suzuki95,Loudon96,Sprik96,Bush&John98,Wang03} This
means that photonic crystals may completely control the emission
rates between complete inhibition and strong enhancement even in the
absence of a photonic bandgap. Since photonic crystals are a natural
platform for solid-state emitters such as quantum
dots,\cite{Lodahl04,Badolato05,Finley05,Englund05} such control of
spontaneous emission is relevant to applications in quantum
information.

Most theoretical papers on spontaneous emission in photonic crystals
concern single light
sources.\cite{Vats02,Sprik96,Suzuki95,Bush&John98,Wang03} In the
case of a weak emitter-field interaction, one expects to see a
single-exponential decay curve with a slope equal to the decay rate
or inverse lifetime. However, experiments on ensembles of emitters
often show nonexponential decays. Such complex decay dynamics can be
due to four reasons: (i) Emitters experience different LDOSs when
they are distributed over different positions and dipole
orientations in the unit cell of a photonic crystal. (ii) It has
been predicted that single sources reveal nonexponential decay due
to van Hove singularities in the LDOS.\cite{Vats02} (iii)
Nonexponential decay may appear if the emitters have more internal
levels than the usually considered two levels. (iv) Temporal
fluctuations of the emitters' environment on time-scales larger than
the fluorescence lifetime can lead to apparent nonexponential
decays. At any rate, it is an open challenge to interpret complex
nonexponential decay curves, in particular, to obtain information on
the local density of states.

In this study, we investigate time-resolved spontaneous emission
from an ensemble of light sources distributed over a well-defined
set of positions in the unit cell. We interpret the emission data
with a continuous distribution of emission rates. This distribution
is identified with the distribution of the LDOS over all positions
\textbf{\sf r}$_\textsf{\scriptsize i}$ sampled by sources with
fixed emission frequency and random dipole orientations. The
distribution width shows a striking sixfold variation with the
varying crystal lattice parameter, in qualitative agreement with
intricate calculations of the three-dimensional (3D) LDOS. This
study opens an avenue to the analysis of time-resolved emission from
ensembles of light sources in complex photonic systems.

\section{Experimental details}

\begin{figure}%[t]
\includegraphics[width=0.75\columnwidth]{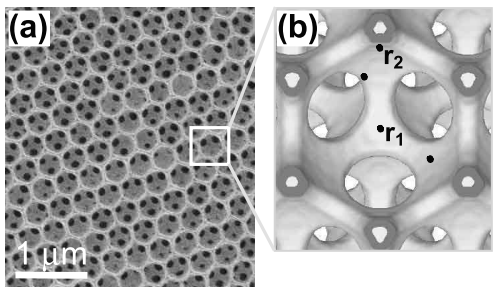}
\caption{\label{sample} (a) Electron-microscope image of the (111)
surface of an inverse opal consisting of air spheres in TiO$_2$ with
lattice parameter \textit{a} = 425 nm. (b) Schematic on quantum dots
(black spots) on the internal titania-air surfaces at
symmetry-inequivalent positions \textbf{\sf r}$_\textsf{\scriptsize
i}$ in the crystal unit cell.}
\end{figure}

We have studied room-temperature spontaneous emission from sources
embedded in fcc inverse opals consisting of air spheres in a titania
(TiO$_2$) backbone shown in Fig.~\ref{sample}(a). These 3D photonic
crystals strongly interact with light.\cite{Koenderink02,Nikolaev05}
Extensive details on the fabrication and characterization of the
inverse opals are reported in Ref.~\onlinecite{Wijnhoven01}. The
lattice parameter \emph{a} is determined from the measured central
wavelength of the lowest stop band $\lambda_c$: $a=
\sqrt{3}\lambda_c/(2n)$, where $n = 1.27 \pm 0.15$ is the average
refractive index consistent with 10 to 20~vol~\% TiO$_2$. We have
studied 15 different samples with lattice parameters ranging from
$\emph{a}$ = 255 $\pm$ 10 nm to $\emph{a}$ = 760 $\pm$ 20 nm.

We use CdSe-ZnSe (core-shell) colloidal quantum dots (QDs) as light
sources because of their high fluorescence quantum efficiency and
narrow homogeneous spectral width.\cite{Dabbousi97} The emission
spectrum of the QDs is centered at $\lambda$ = 610 nm, which is
determined by the average diameter of the nanocrystals of 4.5 nm.
The process of the liquid infiltration of the photonic crystals with
the QDs is described in Refs~\onlinecite{Lodahl04,Nikolaev05}; the
QDs precipitate at positions \textbf{\sf r}$_\textsf{\scriptsize i}$
that are random but within a well defined set, on the internal
surfaces of the air spheres inside the inverse opals (see
Fig.~\ref{sample}b), with an estimated low density of four QDs per
air sphere.

The QDs are excited at $\lambda$ = 447 nm with a diode laser
(Picoquant) emitting 90 ps pulses with 20 nJ/pulse.\cite{Nikolaev05}
We record fluorescence decay curves of QD emission with a
microchannel plate photomultiplier tube detector (Hamamatsu R3809U)
using the time-correlated single-photon counting method. The decay
curves are histograms of the arrival time of a photon emitted after
the laser pulse, obtained over many excitation-emission cycles with
a resolution better than 100 ps. The slope of the decay curves
yields a decay rate $\gamma=\gamma_{rad} + \gamma_{nrad}$, which is
the sum of the spontaneous-emission decay rate $\gamma_{rad}$ and
the nonradiative rate $\gamma_{nrad}$ depopulating the excited
states of the QDs.

\section{Results and discussion}

\begin{figure}
\includegraphics[width=0.75\columnwidth]{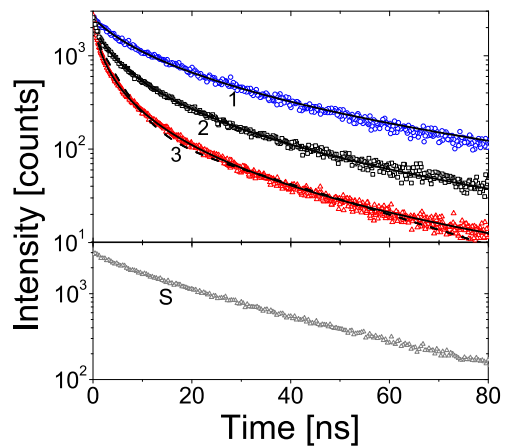}
\caption{\label{decay curves} Fluorescence decay curves recorded at
$\lambda$ = 615 nm and T = 295 K from QDs in inverse opals with
lattice parameters of \textit{a} = 425 nm (1), \textit{a} = 255 nm
(2), and \textit{a} = 540 nm (3); and from QDs in a different
chemical environment, a chloroform suspension (S). The solid lines
are fits of the log-normal distribution of decay rates to the data.
The goodness of fit $\chi^2_{red}$ varies from 1.1 to 1.4, close to
the ideal value of 1. The dashed line in (3) shows a heuristic
biexponential model, which does not agree with the data
($\chi^2_{red}>1.9$).}
\end{figure}

In Fig.~\ref{decay curves} we show time-resolved spontaneous
emission from the QDs in inverse opals with three different lattice
parameters \emph{a}, \emph{i.e.}, different reduced frequencies
$a/\lambda$. The data were collected at $\lambda$ = 615 nm within a
narrow range $\Delta\lambda$ = 3 nm to select the same population of
QDs with identical emission properties on each sample. For inverse
opals with \textit{a} = 255 nm, $\lambda$ = 615 nm is in the
low-frequency limit, where the frequency dependence of the LDOS is
known to show an $\omega^2$-behavior.\cite{Sprik96,Bush&John98} We
see that the spontaneous emission in a sample with \textit{a} = 425
nm is inhibited compared to the reference. Conversely, in a sample
with \textit{a} = 540 nm, the decay rate is enhanced. As in any
photonic-crystal environment, the backbone also fluoresces, which
here distorts the signal at short times. We have carefully removed
the TiO$_2$ signal from the measured decay curves since we know its
spectrum and decay curve from the measurements on an undoped inverse
opal; the backbone has a count rate less than 12~\% of the QD
signal. We exclude the possibility that QDs at the sample surface
contribute any measurable signal: the analysis of separate
angle-resolved measurements reveals an excellent agreement with
theory for sources emitting in the bulk of photonic
crystals.\cite{Nikolaev05} QDs on the surface are rinsed off after
infiltration. Since we also verified that the pump beam was not
Bragg diffracted, the pump intensity is maximum inside the samples
at a distance of several Bragg attenuation lengths from the surface
due to light diffusion.\cite{Nikolaev05} Therefore, Fig.~\ref{decay
curves} demonstrates time-resolved emission from an ensemble of QDs
controlled by photonic crystals, over a much larger time span than
in previous
experiments.\cite{Lodahl04,Badolato05,Finley05,Fujita05,Englund05}

A remarkable feature in Fig.~\ref{decay curves} is that the decay
curves from the QDs in the inverse opals strongly deviate from a
single-exponential decay. To explain this observation, we consider
the four reasons discussed above: i) Since the QDs are distributed
over positions and dipole orientations in the unit cell (see
Fig.~\ref{sample}b), they should indeed experience different LDOSs.
ii) Observation of nonexponential decay due to van Hove
singularities in the LDOS requires single-dot experiments, which is
not the case here. iii) Even though the colloidal QDs are not true
two-level systems, their emission decay is close to being single
exponential,\cite{vDriel05} as confirmed in Fig.~\ref{decay curves}.
iv) It was suggested in Ref.~\onlinecite{Schlegel02Fisher04} that
temporal fluctuations of the environment surrounding the QDs induce
a distribution of nonradiative decay channels. In our experiments,
however, the nonradiative rates $\gamma_{nrad}$ hardly vary from
sample to sample because QDs from the same batch are used, and the
photonic crystals are chemically identical. We observe only minute
differences of decay-curve slopes among the samples with the same
lattice parameters, which indicates that the temporal fluctuations
are identical for all samples. Therefore we can safely attribute the
observed variations of the nonexponential decay curves to a
distribution of radiative decay rates $\gamma_{rad}$ as a result of
a spatial and orientational variation of the LDOS.

To interpret the complex, nonexponential decay curves, we propose a
different line of attack by modeling the curves with a continuous
distribution of decay rates:
\begin{equation}\label{decay-distrib-model}
    I(t)=I(0)\int^{\infty}_{\gamma=0}\phi(\gamma)e^{-\gamma
    t}d\gamma,
\end{equation}
where $\phi(\gamma)$ is a distribution of decay rates with dimension
of time. The fluorescence intensity \emph{I}(t) is proportional to
the time-derivative of the concentration of excited emitters.
Therefore, $\phi(\gamma)$ describes a distribution of the
concentration of emitters with a certain $\gamma$, weighted by the
corresponding $\gamma_{rad}$.\cite{fluorescence-decay} This approach
has two advantages: first, it enables us to explain intrinsically
nonexponential decay curves, and second, the distribution containing
physical information on decay rates is readily available, which is
essential when treating an ensemble of emitters. We use the
log-normal distribution function
\begin{equation}\label{Gaussian-ln-gamma}
\phi(\gamma)=A~\textrm{exp}\biggl(-\frac{\ln^2(\gamma/\gamma_{MF})}{w^2}\biggr),
\end{equation}
where $\gamma_{MF}$ is the most-frequent decay rate corresponding to
the maximum of $\phi(\gamma)$, \emph{w} is a dimensionless width
parameter that determines the distribution width at $1/e$:
\begin{equation}\label{Gaussian-width}
\Delta \gamma = 2 \gamma_{MF} \sinh w .
\end{equation}
\emph{A} is the normalization constant, so that
$\int^{\infty}_{\gamma=0}\phi(\gamma)d\gamma=1$. The important
features of the log-normal distribution are that the logarithmic
form of the distribution function excludes unphysical negative decay
rates and that it is specified in terms of only two free parameters,
$\gamma_{MF}$ and $\Delta\gamma$. Other multiexponential models are
the heuristic biexponential model and the Kohlrausch stretched
exponential model that has been employed to QDs outside photonic
crystals.\cite{Schlegel02Fisher04,Kalkman06} Figure~\ref{decay
curves} shows that the bi-exponential model does not match our data,
even though more free parameters are involved. The stretched
exponential model does not match our data
either,\cite{fluorescence-decay} which again confirms that the
variations we observe are due to LDOS effects in photonic crystals
and not to complex emission properties of the QDs. In
Fig.~\ref{decay curves} it is seen that the log-normal distribution
model (solid curves) provides an excellent description of the
experimental data.

\begin{figure}
\includegraphics[width=0.75\columnwidth]{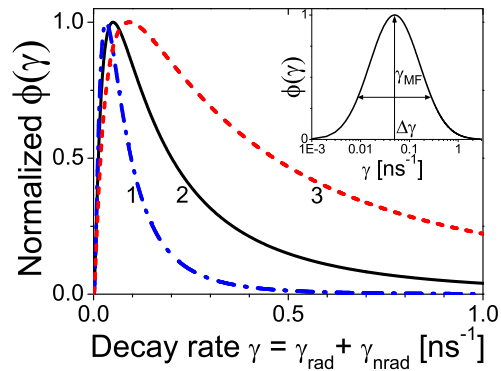}
\caption{\label{distributions} Decay-rate distributions
$\phi(\gamma)$ for the inverse opals with lattice parameters
\textit{a} = 425 nm (1), \textit{a} = 255 nm (2), and \textit{a} =
540 nm (3), corresponding to the data shown in Fig.~\ref{decay
curves}. The inset shows $\phi(\gamma)$ for \textit{a} = 255 nm in
the semilogarithmic scale. Clear modifications of $\Delta\gamma$ and
$\gamma_{MF}$ with varying lattice parameter of the inverse opals
are seen.}
\end{figure}

Figure~\ref{distributions} shows the resulting decay-rate
distributions for three lattice parameters. It is remarkable that
the log-normal distribution model provides an excellent explanation
for all reduced frequencies $a/\lambda$ studied, which will be seen
below to agree with calculations. Compared to the low-frequency
reference (\textit{a} = 255 nm), the maximum of the distribution
$\phi(\gamma)$ is shifted to lower decay rates for the crystal with
\textit{a} = 425 nm and to higher rates for the crystal with
\textit{a}~=~540 nm. These shifts are a clear demonstration of a
photonic effect of the inverse opals on the ensemble of embedded
emitters. In Fig.~\ref{distributions}, we see a dramatic change of
the width $\Delta\gamma$ of the distribution. The large width of
each distribution is identified with the variation of the radiative
emission rates due to orientational and positional \textbf{\sf
r}$_\textsf{\scriptsize i}$ dependences of the LDOS at each lattice
parameter. Consequently, the decay rates of individual QDs are much
more strongly modified by the photonic crystal than the
most-frequent rate $\gamma_{MF}$ of the ensemble.

\begin{figure}%[h]
\includegraphics[width=0.85\columnwidth]{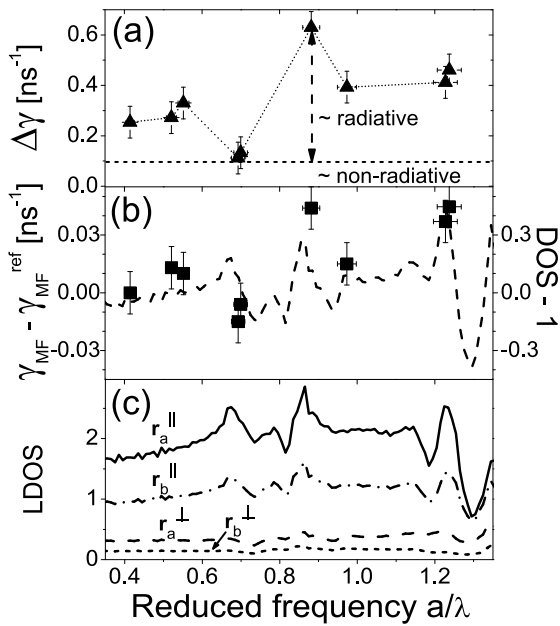}
\caption{\label{results vs red. freq} (a) Width of the decay-rate
distribution $\Delta\gamma$ vs. lattice parameter at a fixed
emission wavelength (triangles, dashed curve is a guide to the eye).
(b) Difference between $\gamma_{MF}$ measured on a photonic sample
and $\gamma_{MF}^{ref}$ = 0.05 ns$^{-1}$ of the low-frequency
reference (squares). The error bars are estimated from the largest
difference between data on the samples with similar~\emph{a}. The
dashed curve represents the relative DOS in the inverse opals. The
relative (L)DOS in the inverse opals is the (L)DOS divided by DOS in
a homogeneous medium with the same average refractive index
(\textit{n} = 1.27). (c) Relative LDOS at two positions on the
internal TiO$_2$-air surfaces projected on dipole orientations
parallel ($\parallel$) or normal ($\perp$) to the internal surface:
at \textbf{\sf r}$_\textsf{\scriptsize a}$ = 0.144(1,1,2) and
\textbf{\sf r}$_\textsf{\scriptsize b}$ = 0.2(1,1,1).}
\end{figure}

In Figs~\ref{results vs red. freq}a and \ref{results vs red. freq}b
we have plotted the resulting values for $\Delta\gamma$ and
$\gamma_{MF}$ versus the reduced frequency. Let us briefly consider
$\gamma_{MF}$: because the nonradiative part of the total decay
rates does not change with the lattice parameter, the change of
$\gamma_{MF}$ - $\gamma_{MF}^{ref}$ is purely radiative and related
to an averaged photonic-crystal LDOS. We compare the experimental
data to the calculated density of states (DOS) -- the unit-cell
average of the LDOS (dashed curve). The measured variation of
$\gamma_{MF}$ is seemingly in good agreement with the DOS. Both
inhibited and enhanced decay rates are observed, and the
experimental variation in $\gamma_{MF}$ amounts to a factor of 3. In
contrast, the width $\Delta\gamma$ shows a striking sixfold
variation, much larger than the change of $\gamma_{MF}$
(Fig.~\ref{results vs red. freq}a). Already in the low-frequency
limit, $a/\lambda$ = 0.4, there is a spatial variation of the
radiative rate ($\Delta\gamma \geq$~0) because the QDs distributed
over inequivalent positions in the unit cell couple to different
electric fields.\cite{Miyazaki98} At the frequencies of the
\emph{L}-gap, $a/\lambda \approx 0.7$, the radiative rate
$\gamma_{rad}$ is inhibited in most places in the unit cell that are
occupied by the QDs, as confirmed by a low continuous-wave (cw)
count rate of only $\thickapprox$ 2.5 kHz. Therefore, the observed
narrow width $\Delta\gamma$ = 0.1 ns$^{-1}$ is a measure of the
distribution width of nonradiative rates. In contrast, at
$a/\lambda$~=~0.88, $\Delta\gamma$ is strongly increased; here the
cw count rate is $\thickapprox$~56~kHz at similar experimental
conditions, in agreement with an enhanced $\gamma_{rad}$. We
therefore conclude that the large widths $\Delta\gamma$ are
determined by a broad distribution of radiative emission rates
$\Delta\gamma_{rad}$ that are proportional to a broad distribution
of the projected LDOS at fixed frequency. Hence, the width
$\Delta\gamma$ is a much more characteristic parameter to describe
the ensemble emission in 3D photonic crystals than the usually used
average rates.\cite{Lodahl04}

We have managed to perform intensive computations of the 3D LDOS at
two representative positions in the unit cell at the TiO$_2$-air
interface (see Fig.~\ref{sample}b). The LDOS shown in
Fig.~\ref{results vs red. freq}c was calculated for dipole
orientations parallel and perpendicular to the TiO$_2$-air
interface. The calculations were performed using 725
reciprocal-lattice vectors in the \emph{H}-field plane-wave
expansion method.\cite{Bush&John98,Wang03} The inverse opals were
modeled as close-packed air spheres surrounded by shells of TiO$_2$
($\epsilon$ = 6.5) with cylindrical windows between neighboring
spheres. The model agrees with prior optical
experiments.\cite{Vos&vDriel00} The integration over wave vectors
\textbf{k} was performed by representing the full Brillouin zone by
an equidistant \textbf{k}-point grid consisting of 291 416
points.\cite{Wang03} The results in Fig.~\ref{results vs red. freq}c
reveal a strong dependence of the LDOS both on the position in the
crystal unit cell (compare curves 1 and 2) and on the dipole
orientation (compare curves 1 and 3). It is remarkable that in the
relevant $a/\lambda$ range, the dependence of the LDOS on the
lattice parameter is the same at both positions \textbf{\sf
r}$_\textsf{\scriptsize i}$ and both orientations, and even as the
unit-cell averaged LDOS (Fig.~\ref{results vs red. freq}b). This
result agrees with the observation that all measured decay curves
are successfully modeled with the same log-normal shape of the
decay-rate distribution. Because the LDOS for dipoles perpendicular
to the interface is inhibited and nearly constant at all reduced
frequencies, whereas the LDOS for parallel dipoles strongly varies,
we propose that the width of the LDOS distribution has a similar
frequency dependence as the LDOS itself. This notion agrees with our
observation that $\Delta\gamma$ tracks the behavior of
$\gamma_{MF}$. A quantitative comparison of our data to the
calculated LDOS is a challenge, since detailed knowledge is needed
on the relation between $\gamma_{rad}$ and $\gamma$ to infer the
true radiative decay-rate distribution (see
Ref.~\onlinecite{fluorescence-decay}). Qualitatively, the calculated
LDOS reflects the main features of our experiments.

\section{Summary}

We have successfully explained highly nonexponential decay curves of
an ensemble of QDs in 3D photonic crystals with a continuous
distribution of decay rates. We relate this distribution to the fact
that QDs in various positions in the unit cell with random dipole
orientations experience different LDOSs. It is gratifying that
recent calculations for Bragg onion resonators also consider
nonexponential decay for similar reasons.\cite{Liang06}

Our results demonstrate that large inhibitions and enhancements of
the spontaneous emission can be achieved with properly positioned
and oriented efficient dipolar light sources inside 3D photonic
crystals, at room temperatures and in large volumes limited only by
the crystal size. The complementary case of a single QD in an opal
has been recently studied in Ref.\onlinecite{Barth06}, and for a
single oriented quantum dot in a two-dimensional slab, interesting
steps have been discussed in
Refs.~\onlinecite{Badolato05,Englund05,Koenderink05}.

\section{Acknowledgments}

We thank L\'{e}on Woldering and Karin Overgaag for experimental
help, Ad Lagendijk and Dani\"{e}l Vanmaekelbergh for stimulating
discussions. This work is a part of the research program of the
Stichting voor Fundamenteel Onderzoek der Materie (FOM) and of
Chemische Wetenschappen (CW) that are financially supported by the
Nederlandse Organisatie voor Wetenschappelijk Onderzoek (NWO).


\begin{thebibliography}{26}

\bibitem{Yab87}
E. Yablonovitch, Phys. Rev. Lett. \textbf{58}, 2059 (1987).

\bibitem{Park04}
H.-G. Park, S.-H. Kim, S.-H. Kwon, Y.-G. Ju, J.-K. Yang, J.-H. Baek,
S.-B. Kim, and Y.-H. Lee, Science, \textbf{305}, 1444 (2004).

\bibitem{Grataetzel01}
M. Gr\"{a}tzel, Nature \textbf{414}, 338 (2001).

\bibitem{Drexhage70} K. H. Drexhage, J. Lumin. \textbf{1–2}, 693 (1970).

\bibitem{Kleppner81} D. Kleppner, Phys. Rev. Lett. \textbf{47}, 233 (1981).

\bibitem{Sprik96}
R. Sprik, B. A. van Tiggelen, and A. Lagendijk, Europhys. Lett.
\textbf{35}, 265 (1996).


\bibitem{Vats02} N. Vats, S. John, and K. Busch, Phys. Rev. A
\textbf{65}, 043808 (2002).

\bibitem{Suzuki95}
T. Suzuki and P.K.L. Yu, J. Opt. Soc. Am. B \textbf{12}, 570 (1995).

\bibitem{Loudon96} S. Barnett and R. Loudon, Phys. Rev. Lett. \textbf{77},
2444 (1996).

\bibitem{Bush&John98} K. Busch and S. John, Phys. Rev. E {\bf 58}, 3896 (1998).

\bibitem{Wang03} R. Wang, X.-H. Wang, B.-Y. Gu, and G.-Z. Yang, Phys. Rev. B \textbf{67},
155114 (2003).

\bibitem{Lodahl04}
P. Lodahl, A.F. van Driel, I.S. Nikolaev, A. Irman, K. Overgaag, D.
Vanmaekelbergh, and W.L. Vos, Nature, \textbf{430}, 654 (2004).

\bibitem{Badolato05} A. Badolato, K. Hennessy, M. Atature, J. Dreiser, E. Hu, P. M. Petroff, and A. Imamoglu, Science \textbf{308}, 1158 (2005).

\bibitem{Finley05} A. Kress, F. Hofbauer, N. Reinelt, M. Kaniber, H. J. Krenner, R. Meyer, G. B\"{o}hm, and J. J. Finley, Phys. Rev. B \textbf{71}, 241304(R) (2005).

\bibitem{Englund05} D. Englund, D. Fattal, E. Waks, G. Solomon, B. Zhang, T. Nakaoka, Y. Arakawa, Y. Yamamoto, and J. Vu\v{c}kovi\'{c}, Phys. Rev. Lett.
\textbf{95}, 013904 (2005).

\bibitem{Koenderink02}
A.F. Koenderink, L. Bechger, H.P. Schriemer, A. Lagendijk, and W.L.
Vos, Phys. Rev. Lett. \textbf{88}, 143903 (2002).

%\bibitem{Ogawa04} S. Ogawa, \emph{et al.}, Science \textbf{305}, 227 (2004).
%

\bibitem{Nikolaev05}
I.S. Nikolaev, P. Lodahl, and W.L. Vos, Phys. Rev. A \textbf{71},
053813 (2005).

\bibitem{Wijnhoven01}
J.E.G.J. Wijnhoven, L. Bechger, and W.L. Vos, Chem. Mater.
\textbf{13}, 4486 (2001).

\bibitem{Dabbousi97} B. O. Dabbousi, J. Rodriguez-Viejo, F. V. Mikulec, J. R.
Heine, H. Mattoussi, R. Ober, K. F. Jensen, and M. G. Bawendi, J.
Phys. Chem. B \textbf{101}, 9463 (1997).

\bibitem{Fujita05} M. Fujita, S. Takahashi, Y. Tanaka, T. Asano, and S. Noda, Science \textbf{308}, 1296 (2005).

\bibitem{vDriel05} A. F. van Driel, G. Allan, C. Delerue, P. Lodahl, W. L. Vos, and D. Vanmaekelbergh, Phys. Rev. Lett.
\textbf{95}, 236804 (2005).

\bibitem{Schlegel02Fisher04}
G. Schlegel, J. Bohnenberger, I. Potapova, and A. Mews, Phys. Rev.
Lett. \textbf{88}, 137401 (2002); B. R. Fisher, H.-J. Eisler, N. E.
Stott, and M. G. Bawendi, J. Phys. Chem. B \textbf{108}, 143 (2004).

\bibitem{fluorescence-decay} A. F. van Driel, I. S. Nikolaev, P. Vergeer, P. Lodahl, D. Vanmaekelbergh, and W. L. Vos, Phys. Rev. B \textbf{75}, 035329 (2007).

\bibitem{Kalkman06} J. Kalkman, H. Gersen, L. Kuipers, and A. Polman, Phys. Rev. B \textbf{73}, 075317
(2006).

\bibitem{Miyazaki98} H. Miyazaki and K. Ohtaka, Phys. Rev. B
\textbf{58}, 6920 (1998).

\bibitem{Vos&vDriel00} W.L. Vos and H.M. van Driel, Phys. Lett. A \textbf{272}, 101
(2000).

\bibitem{Liang06} W. Liang, Y. Huang, A. Yariv, Y. Xu, and S. Y. Lin, Opt. Express
\textbf{14}, 7398 (2006).

\bibitem{Barth06} M. Barth, R. Schuster, A. Gruber, and F. Cichos, Phys. Rev. Lett.
\textbf{96}, 243902 (2006).

\bibitem{Koenderink05} A. F. Koenderink, M. Kafesaki, C. M. Soukoulis, and V. Sandoghdar, Opt. Lett. \textbf{30},
3210 (2005).

\end{thebibliography}
\end{document}